# Excess molar enthalpies of (iodobenzene, or 1-iodonaphthalene + *n*-alkane) liquid mixtures at $T$ = 298.15 K and $p$ = 93 kPa

Fernando Hevia, Luis Felipe Sanz, Juan Antonio González*, Daniel Lozano-Martín, Susana Villa.

GETEF. Departamento de Física Aplicada. Facultad de Ciencias. Universidad de Valladolid. Paseo de Belén, 7, 47011 Valladolid, Spain.

* Corresponding author, e-mail: jagl@uva.es

**Abstract**

Excess molar enthalpies ($H_m^E$) for iodobenzene, or 1-iodonaphthalene + heptane, + decane, + dodecane, or + tetradecane mixtures at 298.15 K and 93 kPa have been measured using a Tian-Calvet micro-calorimeter. The values of $H_m^E$ are positive and indicate that interactions between like molecules are prevalent. In contrast, our previous results on excess molar volumes ($V_m^E$) are negative for the systems $C_6H_5I$ + heptane, or 1-iodonaphthalene + *n*-alkane, which reveal the existence of large structural effects in such solutions. This set of measurements has been used to determine isochoric excess molar internal energies ($U_{V\text{m}}^E$). In the range of *n*-alkanes considered (*n* is the number of C atoms of the alkane), values of $U_{V\text{m}}^E$ at equimolar composition decrease from $n = 7$ to $n = 10$ and then slightly increase for systems with $C_6H_5I$, while decrease slowly for mixtures with 1-iodonaphthalene. These trends fit well with the patterns observed for other alkane mixtures containing cyclic molecules. Dispersive interactions are dominant and those between aromatic molecules with a given halogen atom become stronger when the size of this atom increases due to the corresponding increase of molecular polarizability. The mixtures were studied using the DISQUAC and Flory models. The latter was also applied to *n*-alkane solutions with $C_6H_5F$, or 1-methylnaphthalene. Both theories describe accurately the $H_m^E$ data. In terms of the Flory's model, this means that the random mixing hypothesis is largely achieved. On the other hand, the theory overestimates the interactional contribution to $V_m^E$, particularly for systems with $C_6H_5X$ (X = F, I).

# 1. Introduction

We are engaged in a systematic investigation on mixtures formed by a cyclic molecule A, of plate-like or of more or less globular shape, and an *n*-alkane. Two noticeable effects may exist in the mentioned solutions when long chain *n*-alkanes are involved: Patterson's or Wilhelm's effects [1-4]. The former is typically encountered in binary mixtures with benzene or cyclohexane or $CCl_4$. In such cases, there is an extra endothermic contribution to the excess molar enthalpy ($H_m^E$), which arises from the disruption of correlations of molecular orientations (CMO), a local order characteristic of longer *n*-alkanes [1,5,6]. Wilhelm's effect is encountered, e.g., in *n*-alkane systems involving 1,2,4-trimethylbenzene [7] or 1,2.4-trichlorobenzene [3] or 1-chloronaphthalene [4] (i.e., flat molecules). These solutions show an unexpected behaviour since their values of $H_m^E$ at equimolar composition and 298.15 K decrease when *n*, the number of C atoms of the *n*-alkane, is increased. This has been explained assuming the creation of some type of intramolecular order ascribed to the flat component hinders the rotational motion of the segments of the flexible molecules of longer *n*-alkanes. The application of group contribution methods may be useful to investigate the existence of any of the effects mentioned above [7,8]. When Patterson's effect exists, the theoretical values of $H_m^E$ are lower than the experimental results. If Wilhelm's effect exists, the opposite trend is encountered. Thus, for benzene mixtures, the UNIFAC (Dortmund) results on $H_m^E$/J mol$^{-1}$, obtained using interaction parameters from the literature [9], are: 843 ($n$ = 7); 957 ($n$ = 10); 1052 ($n$ = 14); 1087 ($n$ = 16). The experimental values are (in the same units): 931 ($n$ = 7) [10]; 1031 ($n$ = 10) [11]; 1183 ($n$ = 14) [12]; 1256 ($n$ = 16) [12]. A similar trend is observed when the DISQUAC model [13] is applied (Figure 1). For systems including 1,2,4-trimethylbenzene, UNIFAC provides $H_m^E$/J mol$^{-1}$ = 337 ($n$ = 7); 396 ($n$ = 10); 451 ($n$ = 14); 472 ($n$ = 16). The experimental $H_m^E$/J mol$^{-1}$ results are [7]: 275 ($n$ = 7); 252 ($n$ = 10); 258 ($n$ = 14); 268 ($n$ = 16). These differences between experimental and theoretical results can be only avoided using interaction parameters which are dependent on *n*. In the recent past, we have paid attention to systems of the type $C_6H_5X$ (X = F, Cl, Br), or 1-chloronaphthalene [14,15], or 1,2,4-trichlorobenzene [15] or a bicyclic compound (e.g., bicyclohexyl, cyclohexylbenzene, or decalin) [16] + *n*-alkane. Some relevant conclusions of these studies are the following. (i) Dispersive interactions are dominant. (ii) Large structural effects are present in solutions involving shorter *n*-alkanes (Figure 2). (iii) Excess molar properties at constant volume, internal energies ($U_{Vm}^E$) and heat capacities, are very useful tools to attain a better understanding of interactional and structural effects present in the systems. (iv) Two competing contributions to $U_{Vm}^E$ seem to exist: (a) a poorer ability of longer *n*-alkanes to break interactions between A molecules, which leads to decreased values of $U_{Vm}^E$, and (b) an extra endothermic contribution to this excess function arising from the disruption of CMO of longer $n$-alkanes by the A molecules. If the first contribution dominates, then $U_{V\mathrm{m}}^{\mathrm{E}}$ decreases when $n$ is increased. This is the case of systems with $C_6H_5X$ (X = Cl, Br), 1,2,4-trichlorobenzene, or 1-chloronaphthalene [15]. Solutions involving cyclohexane or benzene show a different behavior: $U_{V\mathrm{m}}^{\mathrm{E}}$ decreases up to $n$ = 8 and from $n \geq 10$ increases, which indicates that, for the solutions with longer $n$-alkanes, the second contribution to $U_{V\mathrm{m}}^{\mathrm{E}}$ is dominant [16]. More recently, we have shown that the determination of $U_{V\mathrm{m}}^{\mathrm{E}}(n)$ is a suitable procedure in order to investigate the possible folding of certain $n$-alkanoates [17], a matter supported by spectroscopic measurements [18,19] and thermophysical data. [20-22]. As a continuation of these works, we provide now experimental $H_{\mathrm{m}}^{\mathrm{E}}$ values for the systems iodobenzene, or 1-iodonaphthalene + heptane, or + decane, or + dodecane, or + tetradecane at 298.15 K and 93 kPa. These new data, together with our previous volumetric measurements for the same mixtures [23,24], are used to determine the corresponding $U_{V\mathrm{m}}^{\mathrm{E}}$ values in order to attain a better characterization of the mentioned solutions. At this end, the systems are also investigated using the DISQUAC [13] and Flory [25] models. Finally, we extend our previous studies on this type of mixtures by applying the latter model to 1-methylnaphthalene or $C_6H_5F$ + $n$-alkane systems.

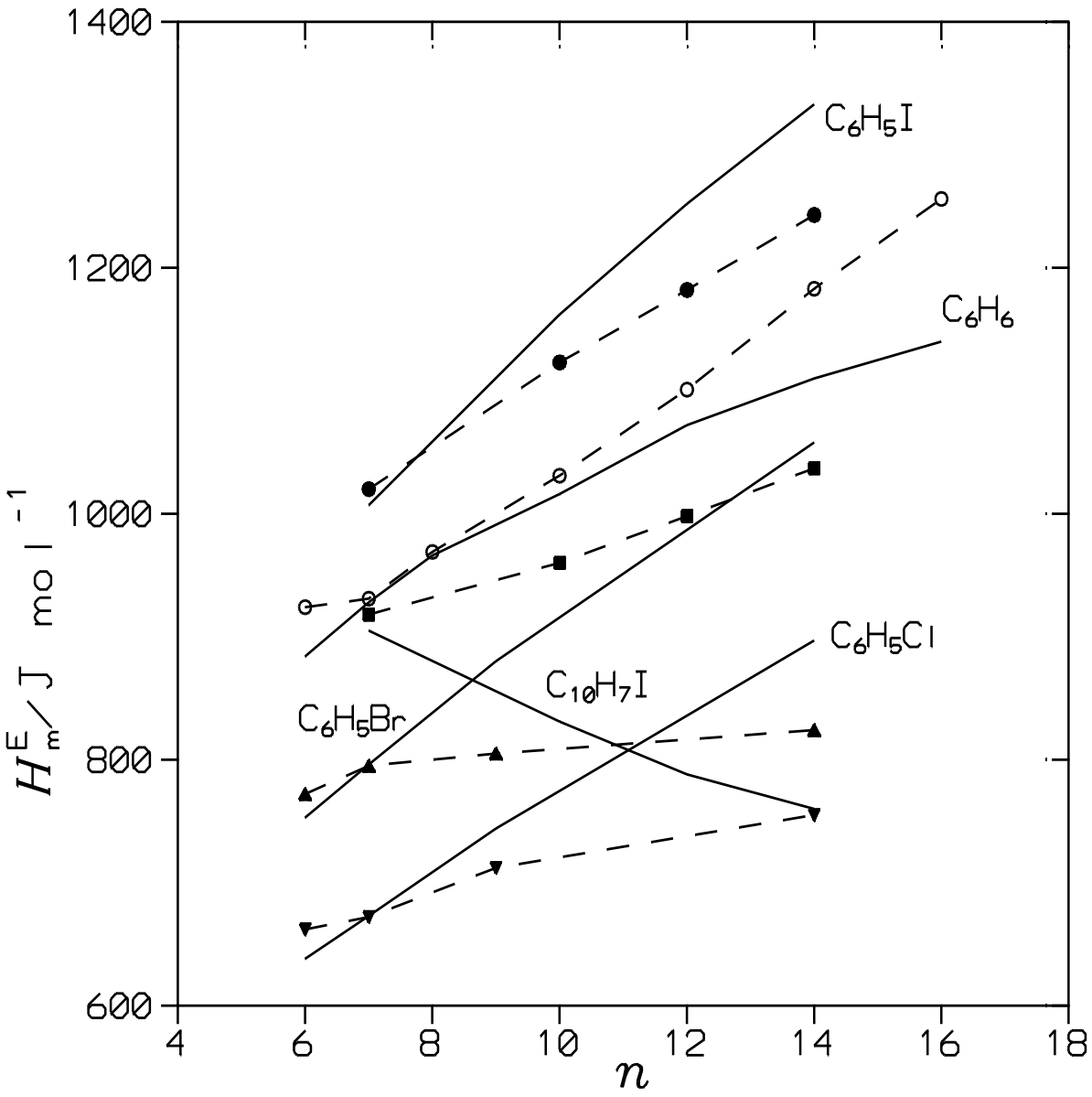

**Figure 1** $H_{\mathrm{m}}^{\mathrm{E}}$ of aromatic compound (1) + $n$-alkane (2) mixtures at equimolar composition, 298.15 K and atmospheric pressure vs. $n$, the number of C atoms in the alkane. Points, experimental results: (●), $C_6H_5I$; (■), 1-iodonaphthalene; (▲), $C_6H_5Br$; (▼), $C_6H_5Cl$; (O), $C_6H_6$. Dashed lines are for the aid of the eye. Solid lines, DISQUAC calculations using interaction parameters for systems with heptane. Source of data: this work and [14,15].

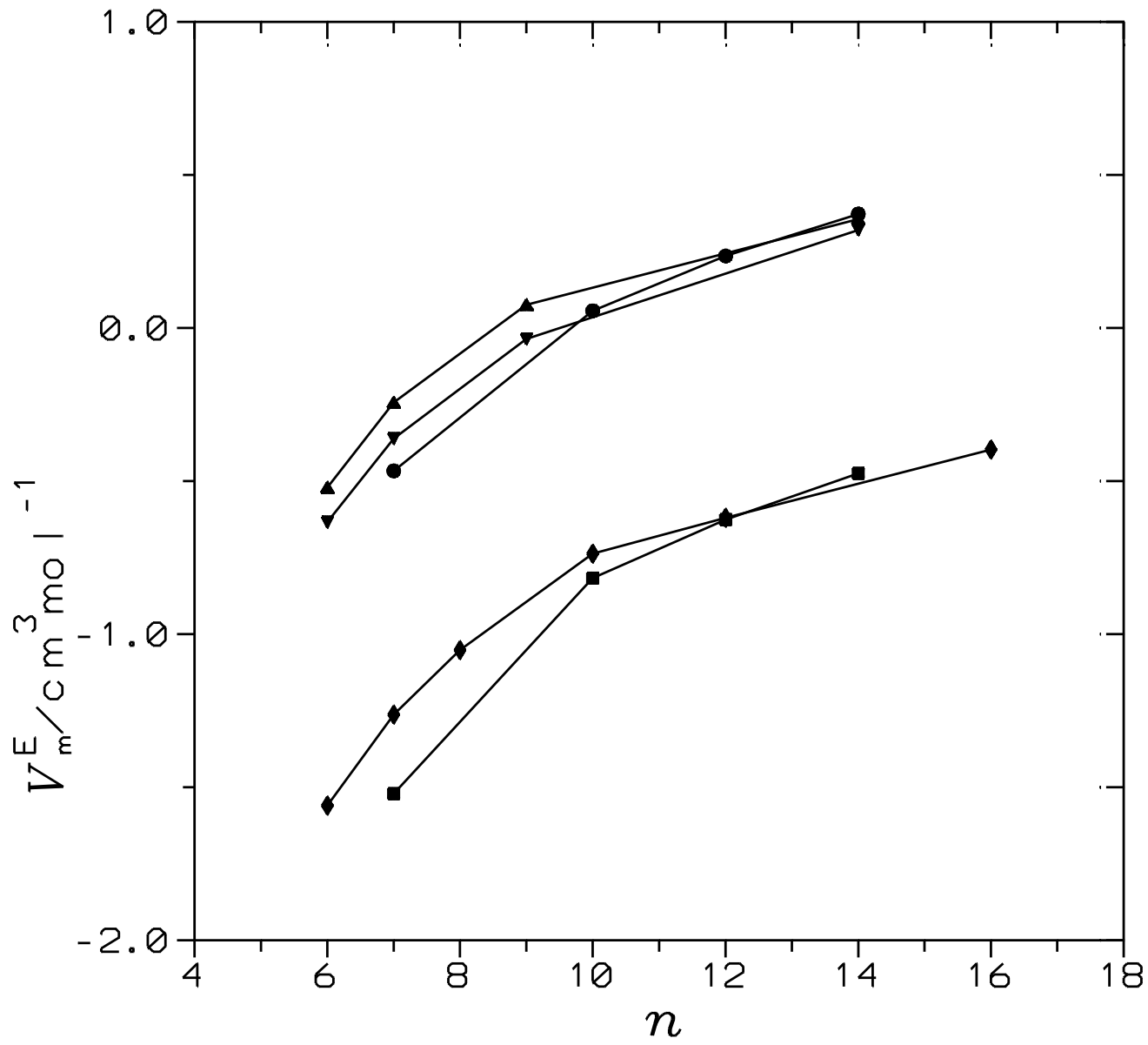

**Figure 2** $V_m^E$ of aromatic halogenated compound (1) + *n*-alkane (2) mixtures at equimolar composition, 298.15 K and atmospheric pressure vs. *n*, the number of C atoms in the alkane. Points, experimental results: (●), $C_6H_5I$; (■), 1-iodonaphthalene; (▲), $C_6H_5Cl$; (▼), $C_6H_5Br$; (◆), 1-chloronaphthalene. Solid lines are for the aid of the eye. For source of data, see [14,15,23,24].

# 2. Experimental

## 2.1. Materials

Pure liquids were used as received from the supplier, without any further purification. Table 1 contains information on the source and purity of the chemicals. Table 1 also shows densities ( $\rho_i$ ) of the pure liquids at 298.15 K and 93 kPa measured using a densitometer Anton Paar DMA 602. The temperature stability of the equipment was 0.01 K, and the repeatability of the density measurements is estimated to be $5\cdot10^{-2}$ kg m$^{-3}$. Details on the calibration of the apparatus can be found elsewhere [26,27]. There is a good agreement between measured and literature data (Table 1).

## 2.2. Apparatus and procedure

Mixtures were prepared by weighing in small vessels of about 10 cm$^3$. The concentration of the mixtures (given by the mole fraction of the iodinated compound, $x_1$) was calculated from mass measurements. The masses were weighted using an analytical balance (MSU125P, Sartorius) and correcting for buoyancy effects. The standard uncertainty of these measurements is $5\times10^{-5}$ g. During the process, caution was taken in order to avoid evaporation. Conversion to molar quantities was based on the relative atomic mass Table of 2015 issued by I.U.P.A.C [28].

The standard uncertainty in the final mole fraction is estimated to be 0.0005. All measurements were performed at 298.15 K and 93 kPa. Excess molar enthalpies, $H_{\mathrm{m}}^{\mathrm{E}}$, were obtained by means of a standard Tian-Calvet micro-calorimeter with a temperature stability of 0.01 K. The mixing cell, designed by us, is of aluminium and has a small gas phase (< 2%). Details on the mixing process and on the calibration of the apparatus have been given previously [29,30]. The estimated maximum relative standard uncertainty for $H_{\mathrm{m}}^{\mathrm{E}}$ is 0.015.

**TABLE 1**

Sample description and densitiy ( $\rho_{\mathrm{i}}$ ) of pure compounds at 298.15 K and 93 kPa.[a]

| Chemical name | CAS Number | Source | Initial purity[b] | $\rho_{\mathrm{i}}$ /g cm$^{-3}$ | |
|---|---|---|---|---|---|
| | | | | Exp. | Lit. |
| iodobenzene | 591-50-4 | Sigma-Aldrich | 0.999 | 1.82209 | 1.82361 [59] |
| 1-iodonaphthalene | 90-14-2 | Sigma-Aldrich | 0.988 | 1.73394 | 1.73510 [60] |
| heptane | 142-82-5 | Sigma-Aldrich | 0.998 | 0.67954 | 0.67960 [24,61] |
| | | | | | 0.67951 [62,63] |
| | | | | | 0.67952 [64] |
| decane | 124-18-5 | Sigma-Aldrich | 0.995 | 0.72612 | 0.7262 [65] |
| | | | | | 0.72607 [62] |
| | | | | | 0.72616 [66] |
| dodecane | 112-40-3 | Sigma-Aldrich | 0.998 | 0.74529 | 0.74534 [24] |
| | | | | | 0.7453 [65] |
| | | | | | 0.74540 [64] |
| tetradecane | 629-59-4 | Fluka | 0.995 | 0.75923 | 0.75929 [24] |
| | | | | | 0.7598 [65] |
| | | | | | 0.75935 [66] |

[a]The uncertainties are: $u(T)$ = 0.01 K; $u(p)$ = 10 kPa. The relative combined expanded uncertainty (0.95 level of confidence) is $U_{\mathrm{rc}}(\rho_{\mathrm{i}}) = 0.002$.

[b]In mole fraction. Initial purity measured by gas chromatography, certified by the supplier.

# 3. Experimental results

Table 2 contains our $H_{\mathrm{m}}^{\mathrm{E}}$ results for $C_6H_5I$, or 1-iodonaphthalene + *n*-alkane mixtures at 298.15 K and 93 kPa pressure vs. $x_1$ (see Figures 3 and 4). The data were adjusted to the equation:

$$H_{\mathrm{m}}^{\mathrm{E}} = x_1(1-x_1)\sum_{\mathrm{i}=0}^{k-1} A_{\mathrm{i}}(2x_1-1)^{\mathrm{i}} \qquad (1)$$

The number of coefficients $k$ used in this equation for each mixture was determined by applying an F-test [31] at the 99.5 % confidence level. The coefficients are listed in Table 3, together with the corresponding standard deviations of $H_{\mathrm{m}}^{\mathrm{E}}$, $\sigma\left(H_{\mathrm{m}}^{\mathrm{E}}\right)$, calculated using the equation:

$$\sigma\left(H_{\mathrm{m}}^{\mathrm{E}}\right)=\left[\frac{1}{N-k}\sum\left(H_{\mathrm{m,exp}}^{\mathrm{E}}-H_{\mathrm{m,calc}}^{\mathrm{E}}\right)^{2}\right]^{1/2} \quad (2)$$

where $N$ is the number of data points.

**TABLE 2**

Excess molar enthalpies, $H_{\mathrm{m}}^{\mathrm{E}}$, of iodobenzene (1) or 1-iodonaphthalene (1) + $n$-alkane (2) mixtures at 298.15 K and at 93 kPa vs. $x_1$, the mole fraction of the aromatic compound.[a]

| $x_1$ | $H_{\mathrm{m}}^{\mathrm{E}}$ /J mol$^{-1}$ | $x_1$ | $H_{\mathrm{m}}^{\mathrm{E}}$ /J mol$^{-1}$ | $x_1$ | $H_{\mathrm{m}}^{\mathrm{E}}$ /J mol$^{-1}$ | $x_1$ | $H_{\mathrm{m}}^{\mathrm{E}}$ /J mol$^{-1}$ |
|---|---|---|---|---|---|---|---|
| | | | iodobenzene(1) + $n$-alkane(2) | | | | |
| | $n$-$C_7$ | | $n$-$C_{10}$ | | $n$-$C_{12}$ | | $n$-$C_{14}$ |
| 0.0563 | 213 | 0.0513 | 189 | 0.0642 | 235 | 0.0585 | 219 |
| 0.0987 | 357 | 0.1065 | 380 | 0.1060 | 382 | 0.1136 | 414 |
| 0.1899 | 610 | 0.1500 | 517 | 0.1940 | 650 | 0.1488 | 524 |
| 0.2943 | 829 | 0.2021 | 664 | 0.3140 | 942 | 0.2033 | 692 |
| 0.3914 | 965 | 0.2942 | 876 | 0.3894 | 1071 | 0.3020 | 945 |
| 0.4908 | 1019 | 0.3881 | 1037 | 0.4815 | 1168 | 0.3963 | 1128 |
| 0.5941 | 995 | 0.4979 | 1123 | 0.4882 | 1175 | 0.4897 | 1233 |
| 0.6978 | 878 | 0.6009 | 1115 | 0.5923 | 1199 | 0.5915 | 1271 |
| 0.7986 | 682 | 0.6952 | 1027 | 0.6899 | 1118 | 0.6954 | 1196 |
| 0.8883 | 427 | 0.7978 | 827 | 0.7884 | 929 | 0.7928 | 1012 |
| 0.9469 | 215 | 0.8493 | 682 | 0.8960 | 571 | 0.8499 | 814 |
| | | 0.8964 | 506 | 0.9475 | 317 | 0.8966 | 621 |
| | | 0.9480 | 279 | | | 0.9475 | 356 |
| | | | 1-iodonaphthalene(1) + $n$-alkane(2) | | | | |
| 0.0616 | 257 | 0.0615 | 226 | 0.0666 | 231 | 0.0908 | 293 |
| 0.1099 | 415 | 0.1106 | 382 | 0.1048 | 351 | 0.1165 | 369 |
| 0.1542 | 525 | 0.1555 | 509 | 0.1447 | 458 | 0.2006 | 598 |
| 0.1925 | 606 | 0.2046 | 636 | 0.2121 | 627 | 0.2755 | 772 |
| 0.2934 | 796 | 0.3005 | 804 | 0.2985 | 797 | 0.3950 | 950 |
| 0.3933 | 889 | 0.3995 | 911 | 0.3973 | 931 | 0.4774 | 1018 |
| 0.4920 | 916 | 0.4944 | 956 | 0.4863 | 990 | 0.5876 | 1044 |
| 0.6001 | 860 | 0.5991 | 929 | 0.5997 | 972 | 0.6987 | 961 |
| 0.6929 | 759 | 0.6989 | 843 | 0.7027 | 886 | 0.8003 | 759 |
| 0.7944 | 568 | 0.7974 | 662 | 0.7970 | 707 | 0.8980 | 463 |
| 0.8941 | 297 | 0.8970 | 367 | 0.8429 | 591 | 0.9498 | 234 |
| | | | | 0.8989 | 406 | | |

[a]The uncertainties are: $u(T)$ = 0.01 K; $u(p)$ = 10 kPa and $u(x_1)$ = 0.0005. The relative combined expanded uncertainty (0.95 level of confidence) is $U_{\mathrm{rc}}\left(H_{\mathrm{m}}^{\mathrm{E}}\right)$ = 0.030.

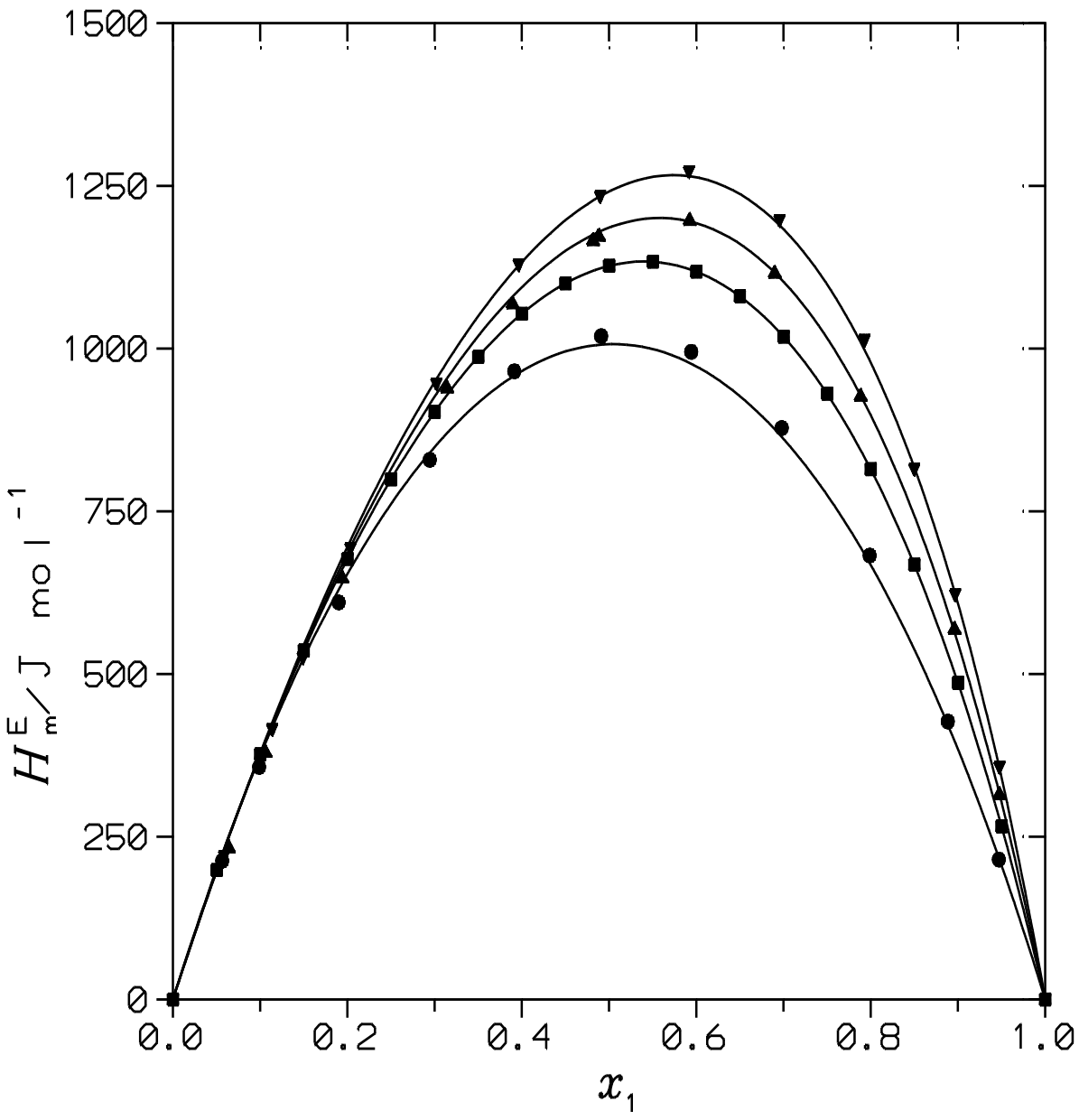

**Figure 3** $H^{\mathrm{E}}_{\mathrm{m}}$ of iodobenzene (1) + *n*-alkane (2) mixtures at 298.15 K and 93 kPa. Points, experimental results (this work): (●), heptane; (■), decane, (▲); dodecane, (▼), tetradecane. Solid lines, DISQUAC calculations using coefficients listed in Table 4.

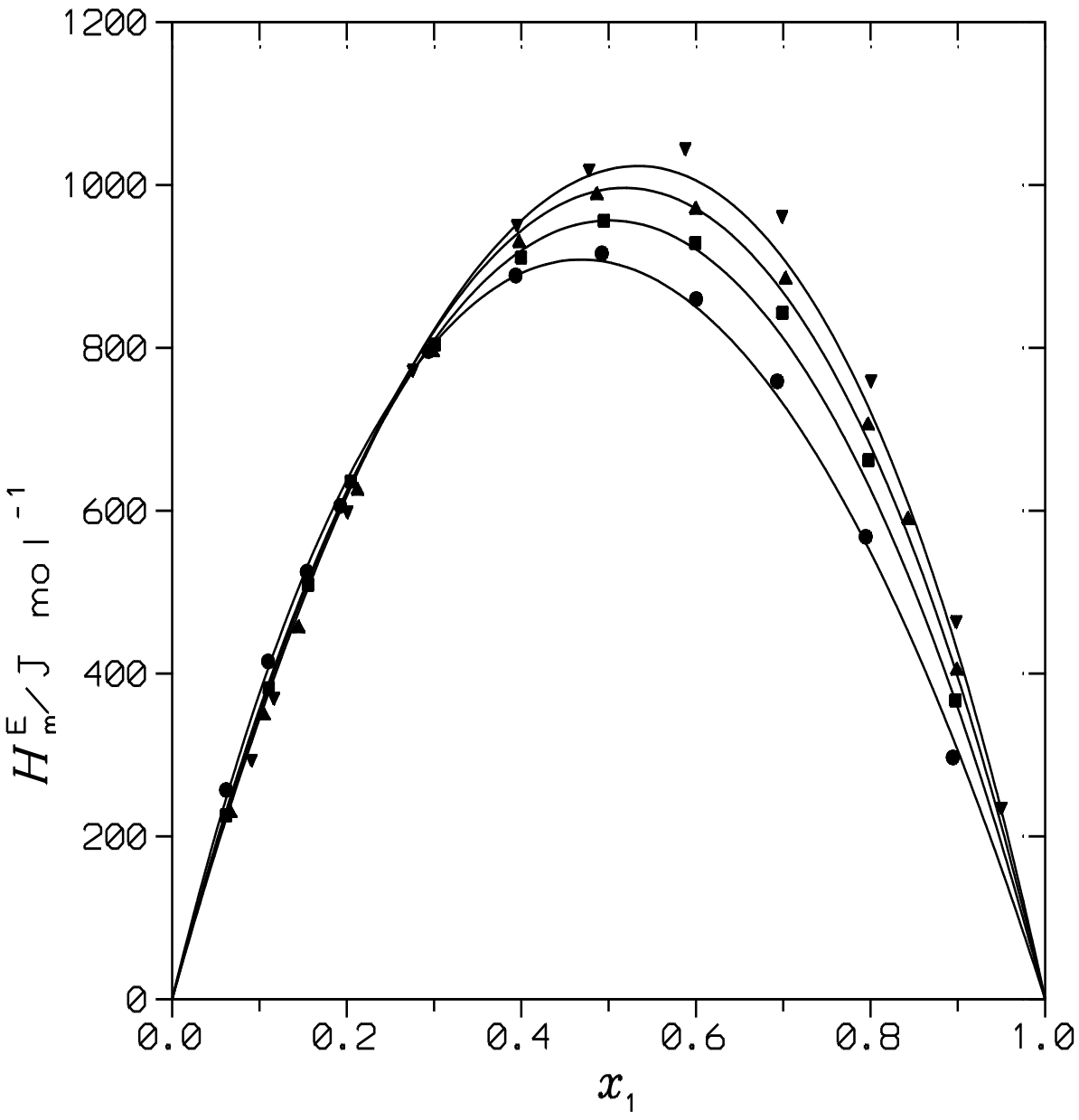

**Figure 4** $H^{\mathrm{E}}_{\mathrm{m}}$ of 1-iodonaphthalene (1) + *n*-alkane (2) mixtures at 298.15 K and 93 kPa. Points, experimental results (this work): (●), heptane; (■), decane; (▲); dodecane, (▼), tetradecane. Solid lines, DISQUAC calculations using coefficients listed in Table 4.

**TABLE 3**

Coefficients $A_i$ and standard deviations, $\sigma(F_m^E)$ (eq. 11) for representation of $F_m^E$ (= $H_m^E$ ; $U_{Vm}^E$) at 298.15 K and 93 kPa for iodobenzene (1) or 1-iodonaphthalene (1) + *n*-alkane (2) systems by eq. (8).

| *n*-alkane | $A_0$ | $A_1$ | $A_2$ | $A_3$ | $\sigma(F_m^E)$ [a]/J mol$^{-1}$ |
|---|---|---|---|---|---|
| | | $F_m^E = H_m^E$ ; iodobenzene (1) + *n*-alkane (2) | | | |
| *n*-$C_7$ | 4093 | 206 | | | 4 |
| *n*-$C_{10}$ | 4494 | 715 | 367 | 362 | 3 |
| *n*-$C_{12}$ | 4726 | 1088 | 510 | 367 | 3 |
| *n*-$C_{14}$ | 4972 | 1431 | 665 | 389 | 4 |
| | | $F_m^E = H_m^E$ ; 1-iodonaphthalene (1) + *n*-alkane (2) | | | |
| *n*-$C_7$ | 3672 | −185 | | −744 | 8 |
| *n*-$C_{10}$ | 3841 | 160 | 313 | | 9 |
| *n*-$C_{12}$ | 3972 | 512 | 241 | | 4 |
| *n*-$C_{14}$ | 4152 | 864 | 241 | | 7 |
| | | $F_m^E = U_{Vm}^E$ ; iodobenzene (1) + *n*-alkane (2) | | | |
| *n*-$C_7$ | 4652 | 327 | | | 2 |
| *n*-$C_{10}$ | 4424 | 912 | 375 | | 6 |
| *n*-$C_{12}$ | 4226 | 1165 | 485 | | 5 |
| *n*-$C_{14}$ | 4486 | 1495 | 624 | | 6 |
| | | $F_m^E = U_{Vm}^E$ ; 1-iodonaphthalene (1) + *n*-alkane (2) | | | |
| *n*-$C_7$ | 5497 | −81 | 568 | −533 | 1 |
| *n*-$C_{10}$ | 4882 | 463 | 681 | | 2 |
| *n*-$C_{12}$ | 4789 | 848 | 411 | | 1 |
| *n*-$C_{14}$ | 4689 | 1178 | 488 | | 2 |

[a] $\sigma(F_m^E) = \left[ \frac{1}{N - k} \sum \left( F_{m,exp}^E - F_{m,calc}^E \right)^2 \right]^{1/2}$

# 4. Models

## 4.1. DISQUAC

DISQUAC is based on the Guggenheim´s rigid lattice theory [32]. Some relevant features of the model are the following. (i) The geometrical parameters (total molecular volumes, $r_i$, surfaces, $q_i$, and molecular surface fractions, $\alpha_i$) of the mixture compounds are calculated additively on the basis of the group volumes $R_G$ and surfaces $Q_G$ recommended by Bondi [33]. The volume $R_{CH4}$ and surface $Q_{CH4}$ of methane are taken arbitrarily as volume and surface units [34]. The geometrical parameters for the considered groups are available in the literature [34-37]. (ii) The partition function is factorized into two terms so that the excess functions are the result of the sum of a dispersive (DIS) term, related to London dispersion forces, and of a quasichemical (QUAC) term, which arises from the anisotropy of the field forces created by the solution molecules. For the molar excess Gibbs energy, $G_m^E$, a combinatorial term, $G_{\mathrm{m}}^{\mathrm{E,COMB}}$, represented by the Flory-Huggins equation [13,34] must be included:

$$G_{\mathrm{m}}^{\mathrm{E}} = G_{\mathrm{m}}^{\mathrm{E,COMB}} + G_{\mathrm{m}}^{\mathrm{E,DIS}} + G_{\mathrm{m}}^{\mathrm{E,QUAC}} \quad (3)$$

$$H_{\mathrm{m}}^{\mathrm{E}} = H_{\mathrm{m}}^{\mathrm{E,DIS}} + H_{\mathrm{m}}^{\mathrm{E,QUAC}} \quad (4)$$

(iii) It is assumed that the interaction parameters depend on the molecular structure. (iv) The coordination number $z = 4$ is used for all the polar contacts. This is a relevant shortcoming of the model and is partially removed using interaction parameters dependent on the molecular structure. (v) It is also assumed that $V_{\mathrm{m}}^{\mathrm{E}}$ (excess molar volume) = 0. The equations required to calculate the DIS and QUAC contributions to $G_{\mathrm{m}}^{\mathrm{E}}$ and $H_{\mathrm{m}}^{\mathrm{E}}$ can be found elsewhere [13,38]. The temperature dependence of the interaction parameters is expressed in terms of the DIS and QUAC interchange coefficients [13,38], $C_{st,l}^{\mathrm{DIS}}, C_{st,l}^{\mathrm{QUAC}}$ where s ≠ t are two contact surfaces present in the mixture and $l = 1$ (Gibbs energy; $C_{st,1}^{\mathrm{DIS,QUAC}} = g_{st}^{\mathrm{DIS,QUAC}}(T_0)/RT_0$); $l = 2$ (enthalpy, $C_{st,2}^{\mathrm{DIS,QUAC}} = h_{st}^{\mathrm{DIS,QUAC}}(T_0)/RT_0$)); $l = 3$ (heat capacity, $C_{st,3}^{\mathrm{DIS,QUAC}} = c_{pst}^{\mathrm{DIS,QUAC}}(T_0)/R$)). $T_0 = 298.15$ K is the scaling temperature and $R$, the gas constant.

## 4.2. Flory model

A summary of the main hypotheses of the theory [25,39,40] can be encountered elsewhere [41] and will not be repeated here. We merely underline that random mixing is a basic assumption of the model. The Flory equation of state is:

$$\frac{\bar{P}\bar{V}}{\bar{T}} = \frac{\bar{V}^{1/3}}{\bar{V}^{1/3} - 1} - \frac{1}{\bar{V}\bar{T}} \quad (5)$$

where $\bar{V} = V_{\mathrm{m}}/V_{\mathrm{m}}^*$; $\bar{P} = P/P^*$ and $\bar{T} = T/T^*$ are the reduced properties: volume, pressure and temperature, respectively. Equation (5) is applied to pure liquids and liquid mixtures. In the former case, the reduction parameters, $V_{\mathrm{mi}}^*$, $P_{\mathrm{i}}^*$ and $T_{\mathrm{i}}^*$ are determined from their data on molar volumes, ($V_{\mathrm{mi}}$), isobaric expansion coefficients ($\alpha_{p\mathrm{i}}$), and isothermal compressibilities ($\kappa_{T\mathrm{i}}$). Expressions for reduction parameters of mixtures are given elsewhere [41]. $H_{\mathrm{m}}^{\mathrm{E}}$ values are determined from:

$$H_{\mathrm{m}}^{\mathrm{E}} = \frac{x_1 V_{\mathrm{m1}}^* \theta_2 X_{12}}{\bar{V}} + x_1 V_{\mathrm{m1}}^* P_1^* \left( \frac{1}{\bar{V}_1} - \frac{1}{\bar{V}} \right) + x_2 V_{\mathrm{m2}}^* P_2^* \left( \frac{1}{\bar{V}_2} - \frac{1}{\bar{V}} \right) \quad (6)$$

In this equation, all the symbols have their usual meaning [41]. The reduced volume of the mixture, $\bar{V}$, in equation (6) is obtained from equation (5). That is, $V_{\mathrm{m}}^{\mathrm{E}}$ can be also calculated:

$$V_{\mathrm{m}}^{\mathrm{E}} = (x_1 V_{\mathrm{m1}}^{*} + x_2 V_{\mathrm{m2}}^{*})(\bar{V} - \varphi_1 \bar{V}_1 - \varphi_2 \bar{V}_2) \quad (7)$$

# 5. Estimation of interaction parameters

## 5.1. DISQUAC

In the framework of DISQUAC, the iodobenzene, or 1-iodonaphthalene + *n*-alkane mixtures are built by three surfaces: (i) type a, aliphatic ($CH_3$, $CH_2$, in *n*-alkanes); (ii) type b, aromatic ($C_6H_5$-, in iodobenzene or $C_{10}H_7$-, in 1–iodonaphthalene); (iii) type g, iodine (in aromatic iodinated compounds). The general procedure applied when estimating the DISQUAC interaction parameters have been given elsewhere [13,38]. The three surfaces generate three contacts: (a,b), (a,g) and (b,g). The (a,b) contacts are described by dispersive interaction parameters only and are known from the study of alkyl-benzene + alkane mixtures for systems with toluene [35], or from the corresponding treatment of 1-methylnaphthalene + *n*-alkane systems [15]. Due to the lack of the required experimental data, and for the sake of simplicity, the interaction parameters for the (b,g) contacts have been neglected, i.e., $C_{\mathrm{bg},l}^{\mathrm{DIS}}$ ($l$ = 1, 2, 3) = 0 for all the systems. The same approach has been applied previously when investigating other aromatic halogenated + *n*-alkane mixtures [15]. Therefore, only the parameters for the (a,g) contacts must be determined (see Table 4). These contacts were represented by DIS and QUAC parameters, since iodobenzene and 1-iodonaphthalene are polar molecules. Their dipole moments are 1.43 D and 1.49 D, respectively [42].

## 5.2. Estimation of the Flory interaction parameter

For $C_6H_5X$ (X = F, I), 1-iodonaphthalene, 1-methylnaphthalene, *n*-heptane, *n*-decane, *n*-dodecane and *n*-tetradecane, values of $V_{\mathrm{mi}}$, $\alpha_{pi}$, $\kappa_{Ti}$ and of reduction parameters at 298.15 K are listed in Table S1. For other alkanes, the corresponding values were taken from previous applications [43]. The interaction parameters, $X_{12}$, (see Table 5) were determined from experimental $H_{\mathrm{m}}^{\mathrm{E}}$ data at equimolar composition and 298.15 K according with the method provided in reference [44].

**TABLE 4**

Dispersive (DIS) and quasichemical (QUAC) interchange coefficients, $C_{\mathrm{ag},l}^{\mathrm{DIS}}$ and $C_{\mathrm{ag},l}^{\mathrm{QUAC}}$, for (a,g) contacts[a] in aromatic iodinated compound (1) + *n*-alkane (2) mixtures ($l$ = 1, Gibbs energy; $l$ = 2, enthalpy; $l$ = 3, heat capacity).

| *n*-alkane | $C_{\mathrm{ag},1}^{\mathrm{DIS}}$ | $C_{\mathrm{ag},2}^{\mathrm{DIS}}$ | $C_{\mathrm{ag},3}^{\mathrm{DIS}}$ | $C_{\mathrm{ag},1}^{\mathrm{QUAC}}$ | $C_{\mathrm{ag},2}^{\mathrm{QUAC}}$ | $C_{\mathrm{ag},3}^{\mathrm{QUAC}}$ |
|---|---|---|---|---|---|---|
| | | | iodobenzene (1) + *n*-alkane (2) | | | |
| *n*-$C_7$ | −1.65 | −0.96 | 0.1[b] | 2 | 1.75 | 0.2 |
| *n*-$C_{10}$ | −1.65 | −1.01 | 0.2[b] | 2 | 1.75 | 0.2 |
| *n*-$C_{12}$ | −1.65 | −1.05 | 0.4[b] | 2 | 1.75 | 0.2 |
| *n*-$C_{14}$ | −1.65 | −1.08 | 0.7[b] | 2 | 1.75 | 0.2 |
| | | | 1-iodonaphthalene (1) + *n*-alkane (2) | | | |
| *n*-$C_7$ | −1.15 | −1.43 | −0.2[b] | 2 | 1.75 | 0.2 |
| *n*-$C_{10}$ | −1.15 | −1.23 | −0.1[b] | 2 | 1.75 | 0.2 |
| *n*-$C_{12}$ | −1.15 | −1.12 | −0.5[b] | 2 | 1.75 | 0.2 |
| *n*-$C_{14}$ | −1.15 | −1.06 | −0.5[b] | 2 | 1.75 | 0.2 |

[a] Type a, $CH_3$, $CH_2$, in *n*-alkanes; type g, I in iodinated compounds.

[b] Estimated values.

**TABLE 5**

Results from the application of the DISQUAC (DQ) and Flory models to aromatic halogenated compound (1) + *n*-alkane (2) mixtures[a] at 298.15 K: number of experimental points ($N$); Flory interaction parameter ($X_{12}$); relative deviation for $H_{\mathrm{m}}^{\mathrm{E}}$ ( dev($H_{\mathrm{m}}^{\mathrm{E}}$) , equation (8)); $H_{\mathrm{m}}^{\mathrm{E}}$ and $V_{\mathrm{m}}^{\mathrm{E}}$ at equimolar composition.

| *n*-alkane | $N$ | $X_{12}$/J cm$^{-3}$ | $H_{\mathrm{m}}^{\mathrm{E}}$ /J mol$^{-1}$ | | dev($H_{\mathrm{m}}^{\mathrm{E}}$) | | | $V_{\mathrm{m}}^{\mathrm{E}}$ /cm$^3$ mol$^{-1}$ | |
|---|---|---|---|---|---|---|---|---|---|
| | | | Exp. | DQ | Exp. | DQ | Flory | Exp. | Flory |
| | | | | iodobenzene (1) + *n*-alkane (2) | | | | | |
| *n*-$C_7$ | 11 | 41.33 | 1023 | 1007 | 0.004 | 0.012 | 0.005 | −0.467 | −0.244 |
| *n*-$C_{10}$ | 13 | 40.48 | 1123 | 1127 | 0.003 | 0.011 | 0.012 | 0.056 | 0.277 |
| *n*-$C_{12}$ | 12 | 40.56 | 1181 | 1186 | 0.002 | 0.008 | 0.013 | 0.234 | 0.460 |
| *n*-$C_{14}$ | 13 | 41.12 | 1241 | 1226 | 0.003 | 0.007 | 0.017 | 0.371 | 0.581 |
| | | | | 1-iodonaphthalene (1) + *n*-alkane (2) | | | | | |
| *n*-$C_7$ | 11 | 32.72 | 918 | 905 | 0.009 | 0.013 | 0.035 | −1.521 | −1.395 |
| *n*-$C_{10}$ | 11 | 29.53 | 960 | 957 | 0.009 | 0.015 | 0.020 | −0.817 | −0.797 |
| *n*-$C_{12}$ | 12 | 28.65 | 993 | 995 | 0.004 | 0.017 | 0.009 | −0.626 | −0.569 |
| *n*-$C_{14}$ | 11 | 28.51 | 1038 | 1019 | 0.007 | 0.025 | 0.013 | −0.475 | −0.404 |
| | | | | fluorobenzene (1) + *n*-alkane (2) | | | | | |
| *n*-$C_6$ | 1515 | 40.26 | 869 | − | 0.003 | − | 0.023 | 0.246 | 0.803 |
| *n*-$C_7$ | 17 | 39.77 | 888 | − | 0.001 | − | 0.022 | 0.431 | 0.937 |
| *n*-$C_8$ | 18 | 41.11 | 941 | − | 0.003 | − | 0.016 | 0.532 | 1.051 |
| *n*-$C_{10}$ | 18 | 43.13 | 1024 | − | 0.001 | − | 0.015 | 0.664 | 1.155 |
| *n*-$C_{12}$ | 18 | 44.91 | 1092 | − | 0.001 | − | 0.010 | 0.739 | 1.239 |
| *n*-$C_{14}$ | 18 | 46.80 | 1156 | − | 0.003 | − | 0.006 | 0.789 | 1.219 |
| *n*-$C_{16}$ | 18 | 47.46 | 1201 | − | 0.005 | − | 0.011 | 0.833 | 1.239 |
| | | | | 1-methylnaphthalene (1) + *n*-alkane (2) | | | | | |
| *n*-$C_7$ | | 27.68 | 759 | | 0.028 | | | | −0.537 |
| *n*-$C_{10}$ | | 22.25 | 702 | | 0.014 | | | | −0.124 |
| *n*-$C_{12}$ | | 19.38 | 649 | | 0.025 | | | | −0.019 |
| *n*-$C_{14}$ | | 17.74 | 622 | | 0.027 | | | | 0.054 |
| *n*-$C_{16}$ | | 16.33 | 597 | | 0.070 | | | | 0.098 |

[a] For DISQUAC results for systems with fluorobenzene, see [14].

## 6. Theoretical results

Results on $H_{\mathrm{m}}^{\mathrm{E}}$ from DISQUAC and the Flory's model are collected in Table 5 (see Figures 3,4). Table 5 also contains relative deviations for $H_{\mathrm{m}}^{\mathrm{E}}$ defined as:

$$\mathrm{dev}(H_{\mathrm{m}}^{\mathrm{E}}) = \left[ \frac{1}{N} \sum \left( \frac{H_{\mathrm{m,exp}}^{\mathrm{E}} - H_{\mathrm{m,calc}}^{\mathrm{E}}}{H_{\mathrm{m,exp}}^{\mathrm{E}}(x_1 = 0.5)} \right)^2 \right]^{1/2} \tag{8}$$

Both models provide very similar $H_{\mathrm{m}}^{\mathrm{E}}$ results. Values of $V_{\mathrm{m}}^{\mathrm{E}}$ using the Flory's model are larger than the experimental ones (Table 5). It is quite clear that the theory overestimates the interactional contribution to this excess function, particularly for systems with fluorobenzene. Nevertheless, the main features of $V_{\mathrm{m}}^{\mathrm{E}}$ are correctly represented (Table 5, Figure 5): the increasing of this excess function with *n* in systems with a given aromatic halogenated compound, and the large negative values which are typical of solutions containing shorter *n*-alkanes for systems with iodobenzene or 1-iodonaphthalene.

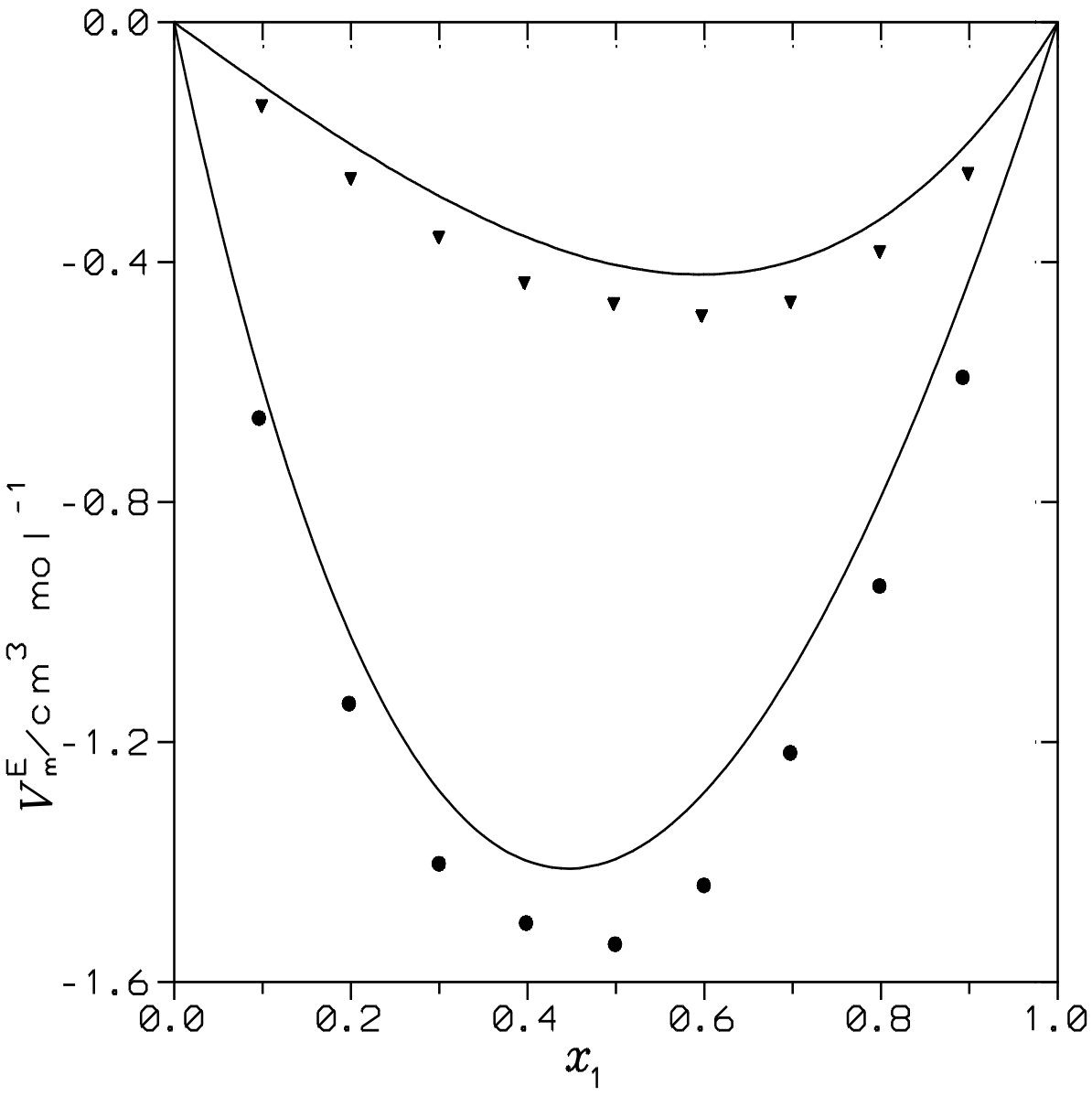

**Figure 5** $V_{\mathrm{m}}^{\mathrm{E}}$ of 1-iodonaphthalene (1) + *n*-alkane (2) mixtures at 298.15 K and 93 kPa. Full points, experimental results [24]: (●), heptane; (▼), tetradecane. Solid lines, results from application of the Flory´s model (Table 5).

# 7. Discussion

Below, the values of the excess molar properties are provided at equimolar composition and 298.15 K.

Iodobenzene or 1-iodonaphthalene + $n$-alkane mixtures show positive values of $H_{\mathrm{m}}^{\mathrm{E}}$ (Table 2, Figures 3, 4), and it is possible to conclude that interactions between like molecules are dominant. In addition, $H_{\mathrm{m}}^{\mathrm{E}}$ values increase with $n$ (Figures 3,4) which is, say, the normal behaviour. The corresponding $V_{\mathrm{m}}^{\mathrm{E}}$ results for the systems with 1-iodonaphthalene or for the solution $C_6H_5I$ + heptane are negative (Figure 2). For example, $V_{\mathrm{m}}^{\mathrm{E}}$(heptane)/cm$^3$ mol$^{-1}$ = $-0.467$ ($C_6H_5I$) [23]; $-1.521$ (1-iodonaphthalene) [24]. This means that, for such systems, $V_{\mathrm{m}}^{\mathrm{E}}$ is mainly determined by structural effects. As previously discussed, the mentioned effects may be of free volume type [23,24]. Compare $\alpha_{pi}/10^{-3}$ K$^{-1}$ = 0.837 ($C_6H_5I$); 0.652 (1-iodonaphthalene) and 1.248 (heptane) (see Table S1). The observed increase of $V_{\mathrm{m}}^{\mathrm{E}}$ with $n$ is, certainly, due to decreasing structural effects [23,24], but also to an increased interactional contribution to $V_{\mathrm{m}}^{\mathrm{E}}$, as it can be deduced from the present enthalpic data. In view of the large structural effects present in a number of the studied solutions, it is quite clear that the equation of state contribution to $H_{\mathrm{m}}^{\mathrm{E}}$ may become very important. For this reason, we have determined $U_{V\mathrm{m}}^{\mathrm{E}}$ from the equation [45,46]:

$$U_{V\mathrm{m}}^{\mathrm{E}} = H_{\mathrm{m}}^{\mathrm{E}} - T\frac{\alpha_p}{\kappa_T}V_{\mathrm{m}}^{\mathrm{E}} \qquad (9)$$

where $\alpha_p$ and $\kappa_T$ are, respectively, the isobaric thermal expansion coefficient and the coefficient of isothermal compressibility of the system under consideration. The $T\frac{\alpha_p}{\kappa_T}V_{\mathrm{m}}^{\mathrm{E}}$ term is the contribution from the Equation of State (EoS) term to $H_{\mathrm{m}}^{\mathrm{E}}$. If the needed data are not available, the $\alpha_p$ and $\kappa_T$ values can be calculated assuming that the mixtures behave ideally with regards to these properties. That is, $M^{\mathrm{id}} = \phi_1 M_1 + \phi_2 M_2$; with $M_{\mathrm{i}} = \alpha_{p\mathrm{i}}$, or $\kappa_{T\mathrm{i}}$ and $\phi_{\mathrm{i}} = x_{\mathrm{i}}V_{\mathrm{mi}}/(x_1 V_{\mathrm{m1}} + x_2 V_{\mathrm{m2}})$. This approximation is commonly acceptable. The values used for $V_{\mathrm{mi}}$, $\alpha_{pi}$ and $\kappa_{T\mathrm{i}}$ are listed in Table S1 (see also references [15] and [43], for other compounds). Results on $U_{V\mathrm{m}}^{\mathrm{E}}$ for mixtures with iodinated compounds are listed in Table 3 (Figures 6 and S1). For solutions with $C_6H_5F$, they are collected in Table S2. No $V_{\mathrm{m}}^{\mathrm{E}}$ data are available in the literature for systems containing 1-methylnaphtahlene which can be used to determine $U_{Vm}^{E}$. It is interesting to underline that for the systems with iodobenzene or 1-iodonaphthalene, the ratio of $(U_{Vm1}^{E,\infty}/U_{Vm2}^{E,\infty})$ ($U_{Vmi}^{E,\infty}$ is the partial excess molar internal energy at constant volume and at infinite dilution of the compound i) changes linearly with $(q_1/q_2)$. Thus, for the mixtures with $C_6H_5I$, $(U_{Vm1}^{E,\infty}/U_{Vm2}^{E,\infty})$ = 0.183 + 0.966 $(q_1/q_2)$, with $r$ = 0.998, and for the solutions with 1-iodonaphthalene, $(U_{Vm1}^{E,\infty}/U_{Vm2}^{E,\infty})$ = $-0.083$ + 1.3665 $(q_1/q_2)$, with $r$ = 0.997. In the case of solutions with $C_6H_5F$ and $n \in [6,14]$, we have: $(U_{Vm1}^{E,\infty}/U_{Vm2}^{E,\infty})$ = 0.270 + 1.067 $(q_1/q_2)$; $r$ = 0.989. This suggests that the change of symmetry of the $U_{Vm}^{E}$ curves is due, at least in part, to size effects. In contrast, linear correlations between $H_{m1}^{E,\infty}/H_{m2}^{E,\infty}$ ($H_{mi}^{E,\infty}$ is the partial excess molar enthalpy at infinite dilution of the compound i) and $(q_1/q_2)$ are poorer for mixtures containing 1-iodonaphthalene: $(H_{m1}^{E,\infty}/H_{m2}^{E,\infty})$ = $-$ 0.588 + 2.332 $(q_1/q_2)$; $r$ = 0.971. Results become improved for systems with $C_6H_5I$, where, noticeably, structural effects are weaker: $(H_{m1}^{E,\infty}/H_{m2}^{E,\infty})$ = 0.036 + 1.207 $(q_1/q_2)$; $r$ = 0.990. This indicates that structural effects are also relevant on the form of the $H_{m}^{E}$ curves.

## 7.1. The effect of the group size

Here, we consider two homologous series: (i) $C_6H_5X$ (X = F, Cl, Br, I) + heptane; (ii) 1-chloro or 1-iodonaphthalene + heptane.

Firstly, we have evaluated the impact of polarity on bulk properties by means of the effective dipole moment $\bar{\mu}_i$ [38,45,47]. For a compound with dipole moment $\mu_i$, $\bar{\mu}_i$ is defined by [38]:

$$\bar{\mu}_i = \left[\frac{\mu^2 N_A}{4\pi\varepsilon_0 V_{mi} k_B T}\right]^{1/2} \tag{10}$$

Here, $N_A$ is Avogadro's constant, $\varepsilon_0$ the permittivity of the vacuum and $k_B$ Boltzmann's constant. For $C_6H_5X$ compounds, $\bar{\mu}_i$ = 0.599 (X = F), 0.598 (X = Cl); 0.585 (X = Br); 0.518 (X = I) (Table S3). These close values suggest that $C_6H_5X$ + heptane mixtures are characterized by similar dipolar interactions. Nevertheless, in previous studies [14,15], we have shown that dispersive interactions are dominant in $C_6H_5X$ + $n$-alkane systems. It may be useful to keep in mind that the excess molar capacity at constant pressure is negative for the heptane solutions with $C_6H_5F$ ($-1.12$ J mol$^{-1}$ K$^{-1}$) [48] or with $C_6H_5Cl$ ($-0.75$ J mol$^{-1}$ K$^{-1}$) [49]. This is a main feature of mixtures characterized by dispersive interactions. The same trend is expected for mixtures with $C_6H_5I$, as it is supported by viscosity data [24,50]. On the other hand, for the first homologous series, $H^{\mathrm{E}}_{\mathrm{m}}$/J mol$^{-1}$ changes in the order: 888 (X = F) [48] > 672 (X = Cl) [51] < 795 (X = Br) [52] < 1023 (X = I), which suggests that the system with $C_6H_5F$ behaves differently. With regards to $U^{\mathrm{E}}_{V\mathrm{m}}$/J mol$^{-1}$, this magnitude varies as follows: 754 (X = F) ≈ 742 (X = Cl) < 901 (X = Br) < 1163 (X = I). It is interesting to compare also the corresponding variations of the $H^{\mathrm{E},\infty}_{\mathrm{m1}}$ and $U^{\mathrm{E},\infty}_{V\mathrm{m1}}$ results. Thus, $H^{\mathrm{E},\infty}_{\mathrm{m1}}$/kJ mol$^{-1}$ = 3.5 (X = F) [48] > 2.9 (X = Cl) [51] < 3.1 (X = Br) [52] < 3.9 (X = I) and $U^{\mathrm{E},\infty}_{V\mathrm{m1}}$/kJ mol$^{-1}$ = 2.9 (X = F) < 3.1 (X = Cl) < 3.3 (X = Br) < 4.3 (X = I). These results remark the importance of the EoS contribution to enthalpic data. It is noteworthy the smooth variation of $U^{\mathrm{E},\infty}_{V\mathrm{m1}}$ for the systems with $C_6H_5X$ (X = F, Cl, Br). On the other hand, the $U^{\mathrm{E},\infty}_{V\mathrm{m1}}$ results show that interactions between $C_6H_5I$ molecules are stronger. The observed increase of $U^{\mathrm{E},\infty}_{V\mathrm{m1}}$ with the size of the halogen atom may be due to the increase of the polarizability of the molecules at this condition. The mean polarizability of a molecule can be estimated from the equation [53,54]:

$$\alpha_{\mathrm{i}} = \frac{3\varepsilon_0 V_{\mathrm{mi}}}{N_{\mathrm{A}}} \frac{n_{\mathrm{Di}}^2 - 1}{n_{\mathrm{Di}}^2 + 2} \quad (11)$$

where $n_{\mathrm{Di}}$ is the refractive index of the compound. Thus, $\alpha_{\mathrm{i}}$/10$^{-24}$ cm$^3$ = 10.3 (X = F) < 12.3 (X = Cl) < 13.4 (X = Br) < 15.5 (X = I) (see Table S3).

In the case of the systems 1-chloronaphthalene or 1-iodonaphthalene + heptane, similar considerations are still valid: dipolar interactions are more or less similar, since the $\bar{\mu}_{\mathrm{i}}$ values are close, 0.514 and 0.471 for 1-chloro and 1-iodonaphthalene, respectively, and dispersive interactions are dominant. Since structural effects are larger in these solutions, large differences are newly encountered between the values of $H^{\mathrm{E}}_{\mathrm{m}}$ and $U^{\mathrm{E}}_{V\mathrm{m}}$, and between the $H^{\mathrm{E},\infty}_{\mathrm{m1}}$ and $U^{\mathrm{E},\infty}_{V\mathrm{m1}}$ results. Thus, $H^{\mathrm{E}}_{\mathrm{m}}$(heptane)/J mol$^{-1}$ = 656 (1-chloronaphthalene) [4]; 918 (1-iodonaphthalene), and, in the same order and units, $U^{\mathrm{E}}_{V\mathrm{m}}$ = 1030 and 1374. For $H^{\mathrm{E},\infty}_{\mathrm{m1}}$(heptane)/kJ mol$^{-1}$ = 3.5 (1-chloronaphthalene) [4]; 4.6 (1-iodonaphthalene), while the corresponding results of $U^{\mathrm{E},\infty}_{V\mathrm{m1}}$ are 4.9 kJ mol$^{-1}$ and 6.7 kJ mol$^{-1}$. Obviously, interactions between 1-iodonaphthelene molecules are stronger. Note that the polarizability of the iodinated compound is also higher ($\alpha_{\mathrm{i}}$/10$^{-24}$ cm$^3$ = 24.5 (1-iodonaphthalene); 20.1 (1-chloronaphthalene) (Table S3).

The existence of dispersive interactions can also be investigated by the situation of the mixtures in the diagram $G^{\mathrm{E}}_{\mathrm{m}}$ vs. $H^{\mathrm{E}}_{\mathrm{m}}$ [55-57]. In this diagram, systems where dipolar interactions are situated between the lines $G^{\mathrm{E}}_{\mathrm{m}}$ = (1/2)$H^{\mathrm{E}}_{\mathrm{m}}$ and $G^{\mathrm{E}}_{\mathrm{m}}$ = $H^{\mathrm{E}}_{\mathrm{m}}$, while those placed between the lines $G^{\mathrm{E}}_{\mathrm{m}} = (1/3)H^{\mathrm{E}}_{\mathrm{m}}$ and $G^{\mathrm{E}}_{\mathrm{m}} = (1/2)H^{\mathrm{E}}_{\mathrm{m}}$ are characterized by dispersive interactions. Here, we consider the situation in a diagram $G^{\mathrm{E}}_{\mathrm{m}}$ vs. $U^{\mathrm{E}}_{V\mathrm{m}}$, since the contribution of the eos term, related to structural effects, to $H^{\mathrm{E}}_{\mathrm{m}}$ may become very large. Note that such effects may be ignored for $G^{\mathrm{E}}_{\mathrm{m}}$ since the corresponding contribution depends on $(V^{\mathrm{E}}_{\mathrm{m}})^2$ and $G^{\mathrm{E}}_{\mathrm{m}} \simeq F^{\mathrm{E}}_{V\mathrm{m}}$ [46]. Using DISQUAC interaction parameters previously determined [14,15] and those listed in Table 4, we have $G^{\mathrm{E}}_{\mathrm{m}}$(heptane)/J mol$^{-1}$ = 435 ($C_6H_5F$); 446 ($C_6H_5Cl$); 479 ($C_6H_5Br$); 510 ($C_6H_5I$); 546 (1-chloronaphthalene); 571 (1-iodonaphthalene), and, in the same order, $G^{\mathrm{E}}_{\mathrm{m}}/U^{\mathrm{E}}_{V\mathrm{m}}$ = 0.57; 0.60; 0.53; 0.44; 0.53; 0.42). That is, dispersive interactions are dominant, particularly for systems with iodinated compounds.

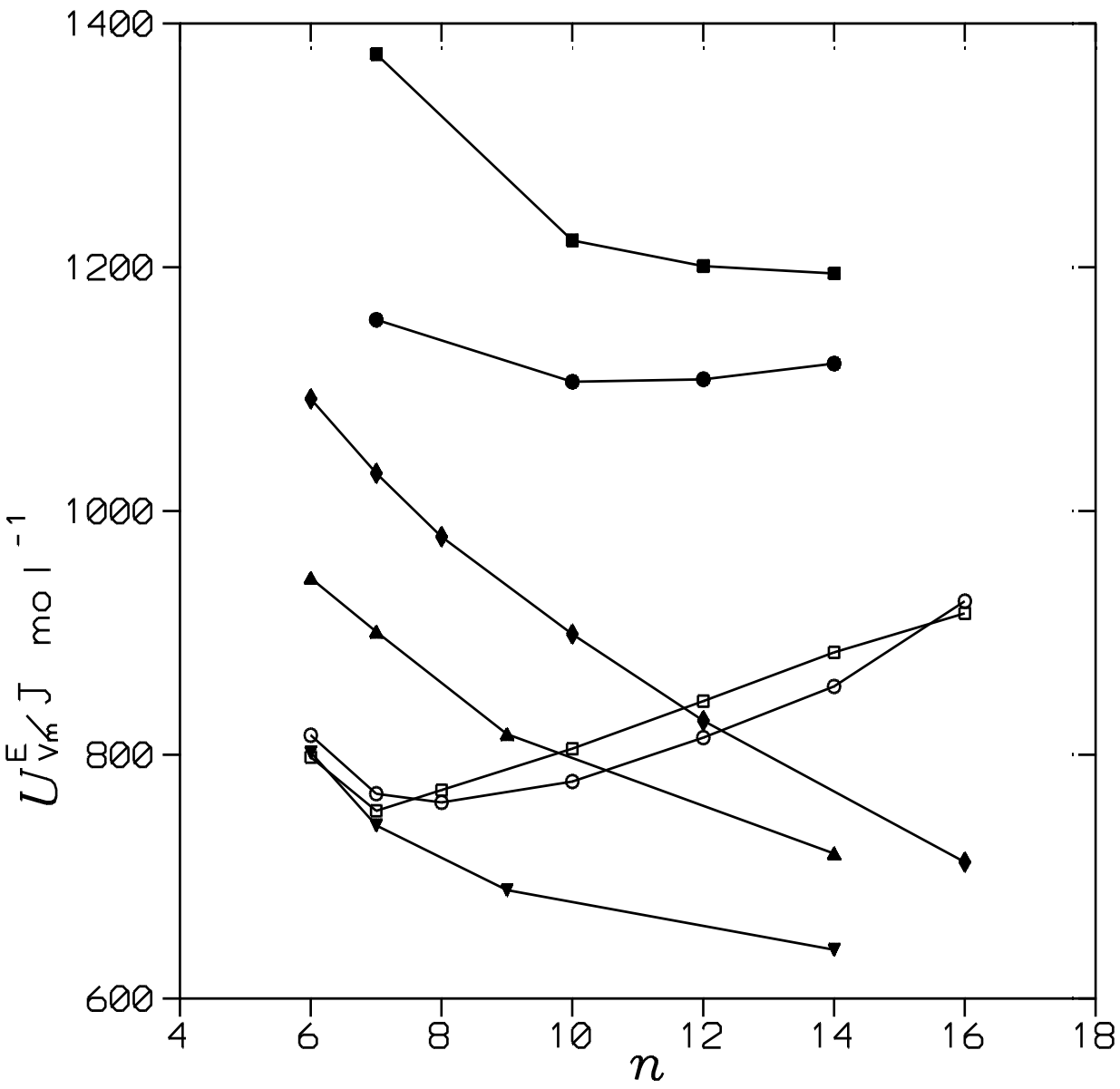

**Figure 6** $U^{\mathrm{E}}_{V\mathrm{m}}$ of aromatic compound (1) + *n*-alkane (2) mixtures at equimolar composition, 298.15 K and atmospheric pressure vs. *n*, the number of C atoms in the alkane. Points, experimental results: (■), 1-iodonaphthalene; (●), $C_6H_5I$; (▲), $C_6H_5Br$; (▼), $C_6H_5Cl$; (□), $C_6H_5F$; (◆), 1-chloronaphthalene; (O), $C_6H_6$. Lines are for the aid of the eye. Source of data: this work and [14,15].

## 7.2. The effect of increasing the aromatic surface (halogen atom fixed)

Inspection of the values reported above indicates that the $H^{\mathrm{E}}_{\mathrm{m}}$ values of systems with $C_6H_5X$ (X = Cl, I) are larger than those of mixtures with 1-chloro or 1-iodonaphthalene (Figure 1). This is due to the large negative contributions to $H^{\mathrm{E}}_{\mathrm{m}}$ from the EoS term for systems with $C_{10}H_7X$ (X = Cl, I). In fact, the $U_{V\mathrm{m}}^{\mathrm{E}}$ values of the systems with 1-chloronaphthalene or 1-iodonaphtalene are larger than for the mixtures with $C_6H_5X$ (X = Cl, I) (Figure 6). Compare, e.g., $U_{V\mathrm{m}}^{\mathrm{E}}$(heptane)/J mol$^{-1}$ = 1157 ($C_6H_5I$), 1374 (1-iodonaphthalene). Furthermore, interactions are stronger between molecules with the larger aromatic surface, which can be explained in terms of the higher polarizability of this type of molecules (Table S3).

## 7.3. Aromaticity effect

Mixtures with 1-iodoalkanes show much lower $H_{\mathrm{m}}^{\mathrm{E}}$ values than those including isomeric aromatic iodinated compounds. Thus, in the case of systems containing 1-iodohexane, $H_{\mathrm{m}}^{\mathrm{E}}$/J mol$^{-1}$ = 436 ($n$ = 6); 601 ($n$ = 12) [36]. The same trend is observed when other halogenated compounds are considered. On the other hand, $H_{\mathrm{m}}^{\mathrm{E}}$ (1-iodoalkane + benzene) < $H_{\mathrm{m}}^{\mathrm{E}}$ (aromatic iodinated compound + heptane). In the former case, interactions between unlike molecules contribute negatively to $H_{\mathrm{m}}^{\mathrm{E}}$, leading to lower values than those of 1-iodoalkane + hexane systems. For example, $H_{\mathrm{m}}^{\mathrm{E}}$(benzene)/J mol$^{-1}$ = 120 (1-iodopentane), 169 (1-iodoheptane) [58]. As usual, the presence of the aromatic surface in the halogenated compound leads to stronger interactions between the polar molecules involved [14].

## 7.4. The variation of $U_{V\mathrm{m}}^{\mathrm{E}}$ with *n*

This variation is typical of *n*-alkane mixtures containing a cyclic molecule. In the range of *n*-alkanes considered, $U_{V\mathrm{m}}^{\mathrm{E}}$ (1-iodonaphthalene) decreases. Mixtures with $C_6H_5X$ (X = Cl, Br) or 1-chloronaphthalene behave similarly, although the variation of $U_{V\mathrm{m}}^{\mathrm{E}}$ with $n$ is sharper (Figure 6). In such cases, the poorer ability of longer *n*-alkanes to break interactions between this type of molecules predominates over the positive contribution to $U_{V\mathrm{m}}^{\mathrm{E}}$ arising from the disruption of CMO of longer *n*-alkanes. $U_{V\mathrm{m}}^{\mathrm{E}}$(iodobenzene) decreases from $n$ = 7 to $n$ = 10 and then smoothly increases (Figure 6), i.e., the two opposite contributions to $U_{V\mathrm{m}}^{\mathrm{E}}$ mentioned above are more or less compensated. For systems with $C_6H_5F$, $U_{V\mathrm{m}}^{\mathrm{E}}$ increases from $n$ = 7, and the contribution from the breaking of the CMO or longer *n*-alkane dominates. Mixtures with benzene behave similarly (Figure 6).

## 7.5. Results from models

The applied models provide good results on $H_{\mathrm{m}}^{\mathrm{E}}$ for the systems studied (Table 5). Results from the Flory's model imply that the random mixing hypothesis is achieved in large extent, and that the systems are mainly characterized by dispersive interactions. Note that the mean deviations for $H_m^E$, defined as $\overline{dev}(H_m^E) = (1/N_S)\sum dev(H_m^E)$ ($N_S$ is the number of systems) are: 0.009 ($C_6H_5I$, $N_S$ = 4; Table 5); 0.015 ($C_6H_5F$, $N_S$ = 7, Table 5): 0.031 ($C_6H_5Cl$, $N_S$ = 4) [15]; 0.025 ($C_6H_5Br$) [15]; 0.032 (1-methylnaphthalene, $N_S$ = 5, Table 5); 0.038 (1-chloronaphthalene; $N_S$ = 6) [15].

With regards to DISQUAC, two considerations must be given: the $C_{ag,l}^{QUAC}$ (l = 1, 2, 3) are the same for the systems with $C_6H_5X$ (X = Cl, Br, I), 1-chloro, or 1-iodonaphthalene. Mixtures with $C_6H_5F$ differ by the $C_{ag,2}^{QUAC}$ value. Note the different $H_m^E$ value for the $C_6H_5F$ + heptane system when compared with those of mixtures involving $C_6H_5X$ (X = Cl, Br) (see above). (ii) If we consider DISQUAC results on $H_m^E$ obtained using $C_{ag,2}^{DIS}$ for heptane solutions, we note that the differences $H_m^E$(DISQUAC) − $H_m^E$(experimental) are positive for systems with $C_6H_5X$ (X = Cl, Br, I) (Figure 1), which suggests the existence of Wilhelm's effect in such solutions. For the sake of comparison, results are also included for benzene systems (Figure 1). In this case, the mentioned differences are negative when longer *n*-alkanes are involved, which points to the existence of Patterson's effect. It has been previously demonstrated that there is no anomalous effect that exists in $C_6H_5F$ + *n*-alkane mixtures [14]. In the case of systems containing 1-iodonaphthalene, DISQUAC predicts decreasing values of $H_m^E$ when *n* is increased, a result which is far from the experimental trend (Figure 1).

# 8. Conclusions

Values of $H_m^E$ for iodobenzene, or 1-iodonaphthalene + heptane, + decane, + dodecane, or + tetradecane mixtures at 298.15 K and 93 kPa have been measured. The systems were studied using the DISQUAC and Flory models. The latter has been also applied to similar solutions with $C_6H_5F$ or 1-methylnaphthalene. The $H_m^E$ values are positive, indicating that interactions between like molecules are prevalent. Large structural effects exist in the systems iodobenzene + heptane, or 1-iodonaphthalene + *n*-alkane. $U_{Vm}^E$ decreases from *n* = 7 to *n* = 10 and then smoothly increases for systems with $C_6H_5I$, while slowly decreases for solutions with 1-iodonaphthalene. The increase of $U_{Vm1}^{E,\infty}$ (heptane) with the size of the halogen atom attached to the aromatic ring is due to the corresponding increase of polarizability, i.e., interactions between aromatic iodinated compounds are stronger. Dispersive interactions are dominant. $H_m^E$ data are well described by DISQUAC and Flory models, although values of $V_m^E$ are overestimated by the latter, particularly for those solutions with $C_6H_5F$ or $C_6H_5I$.

# Funding

This work was supported by Project PID2022-137104NA-I00 funded by MCIN/AEI/10.13039/501100011033/ and by FEDER, UE.

# Supplementary material

# Excess molar enthalpies of (iodobenzene, or 1-iodonaphthalene + *n*-alkane) liquid mixtures at $T$ = 298.15 K and 93 kPa

Fernando Hevia, Luis Felipe Sanz, Juan Antonio González*, Daniel Lozano-Martín, Susana Villa

GETEF. Departamento de Física Aplicada. Facultad de Ciencias. Universidad de Valladolid. Paseo de Belén, 7, 47011 Valladolid, Spain

* Corresponding author, e-mail: jagl@uva.es

**TABLE S1**

Isobaric expansion coefficients ( $\alpha_{pi}$ ), isothermal compressibilities ( $\kappa_{Ti}$ ), molar volumes and reduction parameters of pure compounds at 298.15 K and atmospheric pressure used in Flory calculations.

| Compound | $\alpha_{pi}$ /10$^{-3}$ K$^{-1}$ | $\kappa_{Ti}$ /10$^{-12}$ Pa$^{-1}$ | $V_{mi}$ /cm$^3$ mol$^{-1}$ | $V^*_{mi}$ /cm$^3$ mol$^{-1}$ | $P^*_i$ /J cm$^{-3}$ |
|---|---|---|---|---|---|
| $C_6H_5I$ | 0.837 [s1] | 600.6 [s1] | 111.96 [s1] | 92.28 | 611.2 |
| 1-iodonaphthalene | 0.652 [s2] | 430.2 [s3] | 146.53 [s2] | 125.05 | 620.4 |
| $C_6H_5F$ | 1.18 [s4] | 720 [s4,s5] | 94.31 [s4] | 73.48 | 587.2 |
| 1-methylnaphthalene | 0.693 [s6] | 574 [s6,s7] | 139.86 [s6] | 118.20 | 508.7 |
| heptane | 1.248 [s2] | 1454.2 [s1] | 147.45 [s2] | 113.74 | 430.1 |
| decane | 1.040 [s2] | 1103.9 [s1] | 195.95 [s2] | 156.03 | 442.9 |
| dodecane | 0.971 [s2] | 991.0 [s1] | 228.55 [s2] | 184.06 | 450.4 |
| tetradecane | 0.929 [s2] | 918.4 [s1] | 261.30 [s2] | 211.90 | 458.7 |

**TABLE S2**

Coefficients $A_i$ and standard deviations, $\sigma(U^E_{Vm})$ (eq. 2) for representation of $U^E_{Vm}$ at temperature $T$ and 93 kPa for fluorobenzene (1) + $n$-alkane (2) systems by eq. (9).

| $n$-alkane | $A_0$ | $A_1$ | $A_2$ | $A_3$ | $\sigma(U^E_{Vm})$ /J mol$^{-1}$ |
|---|---|---|---|---|---|
| $n$-$C_6$ | 3192 | 103 | | | 0.3 |
| $n$-$C_7$ | 3016 | 144 | 64 | | 0.5 |
| $n$-$C_8$ | 3086 | 340 | | | 3 |
| $n$-$C_{10}$ | 3219 | 501 | 55 | | 1 |
| $n$-$C_{12}$ | 3377 | 750 | 132 | | 3 |
| $n$-$C_{14}$ | 3536 | 963 | 494 | | 1 |
| $n$-$C_{16}$ | 3663 | 1106 | 480 | 573 | 1 |

**TABLE S3.** Physical properties of pure compounds: molar volumes, $V_{mi}$, refractive indices, $n_D$, mean polarizabilities, $\alpha_i$ , dipole moments, $\mu_i$ and effective dipole moments, $\bar{\mu}_i$ .

| Compound | $V_{mi}{}^a$/cm$^3$ mol$^{-1}$ | $n_D{}^a$ | $\alpha{}^a$/10$^{-24}$ cm$^3$ | $V_{mi}{}^b$/cm$^3$ mol$^{-1}$ | $\mu_i{}^b$/D | $\bar{\mu}_i$ |
|---|---|---|---|---|---|---|
| $C_6H_5F$ | 93.79 [s8] | 1.466 [s8] | 10.3 | 94.31 [s4] | 1.52 [s9] | 0.599 |
| $C_6H_5Cl$ | 101.71 [s8] | 1.525 [s8] | 12.3 | 102.24 [s10] | 1.58 [s9] | 0.598 |
| $C_6H_5Br$ | 104.96 [s8] | 1.560 [s8] | 13.4 | 105.50 [s10] | 1.57 [s9] | 0.585 |
| $C_6H_5I$ | 111.38 [s1] | 1.620 [s8] | 15.5 | 111.96 [s1] | 1.43 [s9] | 0.517 |
| 1-chloro naphthalene | 142.32 [s8] | 1.632 [s8] | 20.1 | 136.77 [s11] | 1.52 [s9] | 0.497 |
| 1-iodo naphthalene | 146.05 [s2] | 1.703 [s12] | 24.5 | 146.53 [s2] | 1.49 [s19] | 0.471 |

[a] Values at 293.15 K; [b] values at 298.15 K.

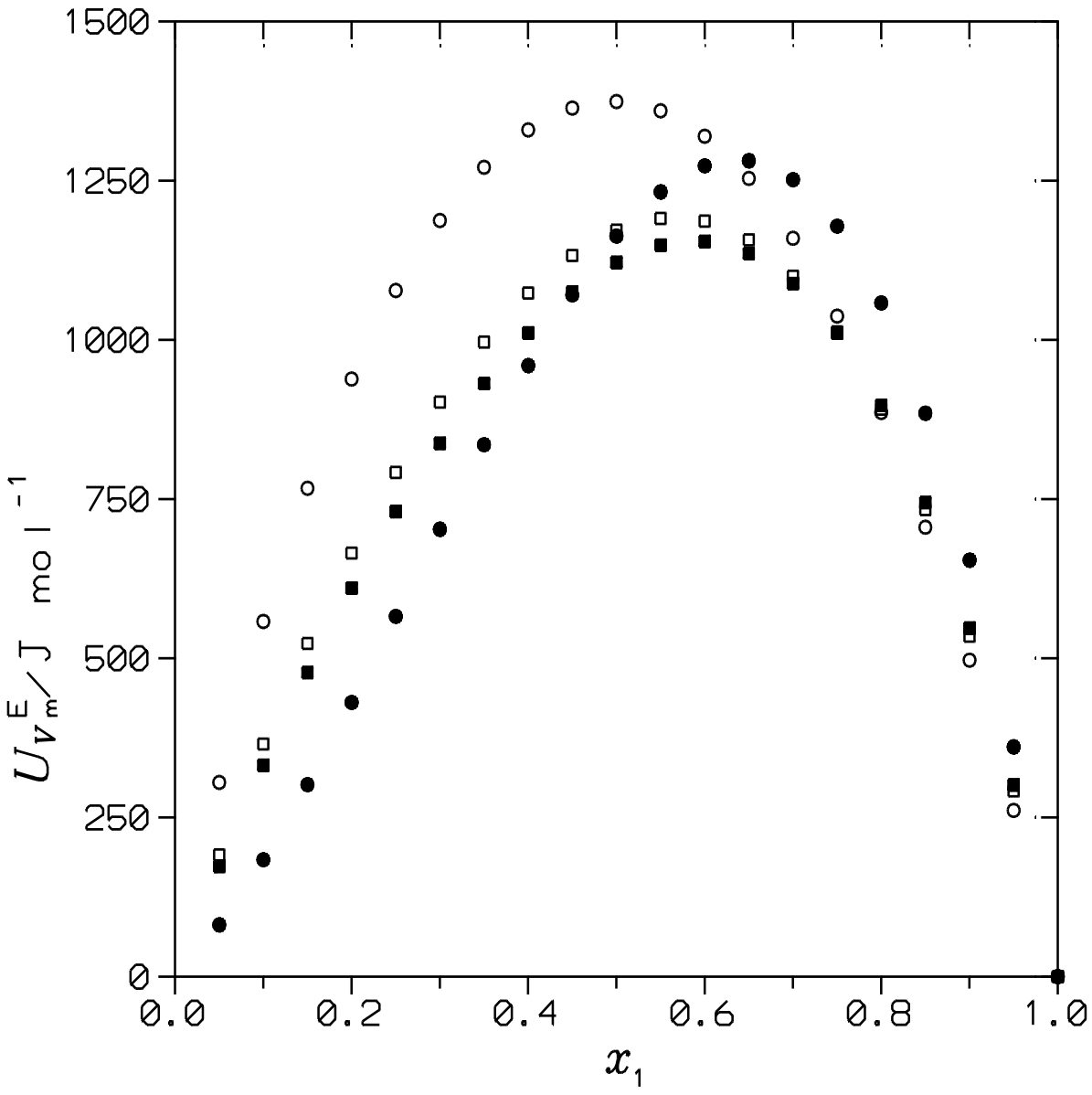

**Figure S1.** $U_{V\mathrm{m}}^{\mathrm{E}}$ of iodobenzene (1) or 1-iodonaphthalene (1) + *n*-alkane (2) mixtures at 298.15 K and 93 kPa. Full points, experimental results for $C_6H_5I$ systems (this work): (●), heptane; (■), tetradecane; open points, mixtures with 1-iodonaphthalene (this work); (O), heptane; (□), tetradecane.